\documentclass[twocolumn,preprintnumbers,prb,aps]{revtex4}
\usepackage{graphicx}
\usepackage{amsmath}
\usepackage{color}
\newcommand{\be}{\begin{equation}}
\newcommand{\ee}{\end{equation}}
\newcommand{\bea}{\begin{eqnarray}}
\newcommand{\eea}{\end{eqnarray}}
\newcommand{\bs}{\begin{split}}

\begin{document}

\title{Magnetic Susceptibility of Collinear and Noncollinear Heisenberg Antiferromagnets}
\author {D. C. Johnston} 
\altaffiliation{johnston@ameslab.gov}
\affiliation {Ames Laboratory and Department of Physics and Astronomy, Iowa State University, Ames, Iowa 50011, USA}

\date{\today}

\begin{abstract}

Predictions of the anisotropic magnetic susceptibility $\chi$ below the antiferromagnetic (AFM) ordering temperatures $T_{\rm N}$ of local moment Heisenberg AFMs have been made previously using molecular field theory (MFT) but are very limited in their applicability.  Here a MFT calculation of $\chi(T\leq T_{\rm N})$ is presented for a wide variety of collinear and noncollinear Heisenberg AFMs containing identical crystallographically equivalent spins without recourse to magnetic sublattices.  The results are expressed in terms of directly measurable experimental parameters and are fitted with no adjustable parameters to experimental $\chi(T\leq T_{\rm N})$ data from the literature for several collinear and noncollinear AFMs.  The influence of spin correlations and fluctuations beyond MFT is quantified by the deviation of the theory from the data.  The origin of the universal $\chi(T\leq T_{\rm N})$ observed for triangular lattice AFMs exhibiting coplanar noncollinear 120$^\circ$ AFM ordering is clarified.

\end{abstract}


\maketitle

\paragraph*{Introduction.} Magnetic susceptibility $\chi$ measurements versus temperature $T$ have been used for a century to obtain important information about the magnetic properties of materials.  The Weiss molecular field theory (MFT) has been instrumental in interpreting the $\chi(T)$ data in the paramagnetic state above the long-range magnetic ordering temperature $T_{\rm N}$ of local magnetic moment antiferromagnets\cite{Johnston2011,Keffer1966} (AFMs) via the Curie-Weiss (CW) law $\chi = \frac{C}{T-\theta_{\rm p}}$, in which the magnitude of the local moments is contained in the Curie constant $C$ and the nature and strengths of their interactions in the Weiss temperature $\theta_{\rm p}$.  MFT has also been used extensively for comparisons with experimental data of its predictions for the ordered magnetic moment and magnetic heat capacity versus $T$ in the ordered state of AFMs at $T < T_{\rm N}$.  Thus MFT  is a primary tool to identify important characteristics of local moment AFMs.  

In contrast, very few comparisons have been made of experimental anisotropic $\chi(T < T_{\rm N})$ data for AFMs  with the predictions of MFT even for collinear AFMs where the ordered moments $\vec{\mu}_i$ are aligned along the same easy axis.\cite{Johnston2011, Singer1956, Keffer1966}  Here we provide simple MFT expressions to fit experimental $\chi(T < T_{\rm N})$ data for ordered AFMs containing identical crystallographically equivalent spins interacting by Heisenberg exchange for arbitrary sets of exchange constants.  The theory treats collinear and planar noncollinear AFM structures on the same footing without the use of magnetic sublattices.  The results are expressed in terms of independent experimentally measurable quantities and are used to fit with no adjustable parameters representative experimental $\chi(T < T_{\rm N})$ data from the literature for several collinear and noncollinear AFMs.  The fits can quantify the influence of spin correlations and fluctuations beyond MFT on $\chi(T < T_{\rm N})$, and can also help to elucidate the AFM structures and exchange interactions if these are uncertain or unknown.

\begin{figure}
\includegraphics [width=2.25in]{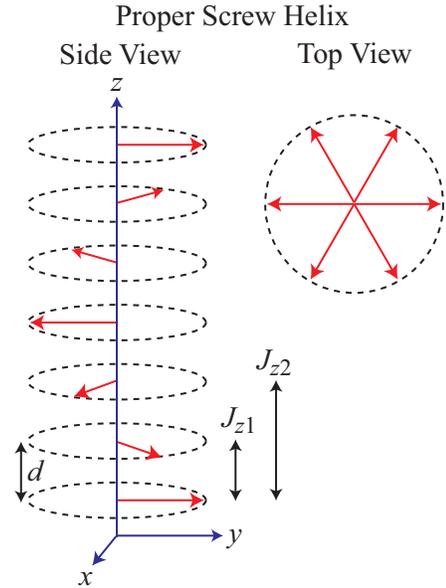}
\caption{(Color online) Yoshimori's ``proper screw helix'' noncollinear AFM structure that he proposed for MnO$_2$.\cite{Yoshimori1959}  Each magnetic moment represents a planar array of ferromagnetically aligned magnetic moments that lie in the $xy$-plane, where the planes are separated along the helix axis ($z$-axis) by distance $d$.  The plane of the magnetic moments is perpendicular to the helix wave vector ${\bf k}$ along the $z$-axis.  The net interplanar exchange interactions $J_{z1}$ and $J_{z2}$ in the generic $J_0$-$J_{z1}$-$J_{z2}$ model are indicated.  A ``cycloidal helix'' AFM structure\cite{Yoshimori1959} occurs when {\bf k} is in the $xy$-plane of the magnetic moments. }
\label{Fig:J0_Jz1_Jz2_model_helix}
\end{figure}

Using MFT, Van~Vleck calculated in 1941 the anisotropic $\chi(T\leq T_{\rm N})$ for magnetic fields {\bf H} applied parallel ($\chi_\parallel$) and perpendicular $(\chi_\perp)$ to the easy axis of collinear AFMs with only nearest-neighbor Heisenberg interactions between spins on two distinct interpenetrating ``bipartite'' sublattices.\cite{VanVleck1941}  Yoshimori carried out MFT calculations of $\chi(T\leq T_{\rm N})$ in 1959 for the special  case of a planar noncollinear AFM ``proper screw helix'' magnetic structure that he proposed for MnO$_2$,\cite{Yoshimori1959, Nagamiya1967} as shown schematically in Fig.~\ref{Fig:J0_Jz1_Jz2_model_helix}.  These MFTs are very restricted in their applicability and have been rarely used to fit experimental $\chi(T\leq T_{\rm N})$ data over the past five decades.

\paragraph*{Theory.} Here we consider identical crystallographically equivalent spins interacting by Heisenberg exchange, with no anisotropy present except that due to an infinitesimal {\bf H}\@. The part $E_i$ of the average energy of the system that is associated with interactions of $\vec{\mu}_i$ with {\bf H} and with its neighbors $\vec{\mu}_j$ is \mbox{$E_i = \frac{1}{2g^2\mu_{\rm B}^2}\vec{\mu}_i\cdot\sum_j J_{ij}\vec{\mu}_j - \vec{\mu}_i \cdot{\bf H}$,} where $J_{ij}$ is the Heisenberg exchange coupling between ordered magnetic moments $\vec{\mu}_i$ and $\vec{\mu}_j$.  Using MFT,\cite{Johnston2011} one obtains the CW law for $T\geq T_{\rm N}$, where $C = \frac{Ng^2\mu_{\rm B}^2S(S+1)}{3k_{\rm B}}$, $\theta_{\rm p} = -\frac{S(S+1)}{3k_{\rm B}}\sum_j J_{ij}$ and $T_{\rm N} = -\frac{S(S+1)}{3k_{\rm B}}\sum_j J_{ij}\cos\phi_{ji}$, $N$ is the number of spins, $g$ is the $g$-factor, $\mu_{\rm B}$ is the Bohr magneton, $S$ is the spin, $k_{\rm B}$ is Boltzmann's constant and $\phi_{ji}$ are the angles between $\vec{\mu}_i$ and its neighbors $\vec{\mu}_j$ in the AFM-ordered state.  We rewrite the CW law for $T\geq T_{\rm N}$ in dimensionless form as
\be
\frac{\chi(t)T_{\rm N}}{C} = \frac{1}{t-f},\quad t\equiv\frac{T}{T_{\rm N}},\quad f\equiv\frac{\theta_{\rm p}}{T_{\rm N}} = \frac{\sum_j J_{ij}}{\sum_j J_{ij}\cos\phi_{ji}}.
\label{Eq:CThetap}
\ee
Below $T_{\rm N}$, the $\chi$ with {\bf H} perpendicular to the ordered moment axis or plane for collinear or planar noncollinar AFMs, respectively, is given in general by MFT as\cite{Johnston2011}
\be
\frac{\chi_\perp(T\leq T_{\rm N}) T_{\rm N}}{C} = \frac{\chi(T_{\rm N}) T_{\rm N}}{C} = \frac{1}{1-f}. \qquad(t\leq1)
\label{Eq:ChiPerp}
\ee

For collinear AFMs, a field applied below $T_{\rm N}$ along the easy axis just changes the magnitude of an  ordered moment without rotating it and in MFT we obtain
\be
\frac{\chi_\parallel(t)T_{\rm N}}{C} = \frac{1}{\tau^\ast-f}, \qquad \tau^\ast(t) = \frac{(S+1)t}{3B_S^\prime(y_0)},
\label{Eq:ChiParaColl}
\ee
where $B_S^\prime(y_0) \equiv dB_S(y)/dy|_{y=y_0}$, $B_S(y)$ is the Brillouin function,\cite{Johnston2011}
$y_0 =\frac{3\bar{\mu}_0}{(S+1)t},\ \bar{\mu}_0 = \frac{\mu_0}{\mu_{\rm sat}},\ \mu_{\rm sat} = gS\mu_{\rm B}$,
and the magnitude of the ordered moment in zero field $\bar{\mu}_0(t)$ is calculated numerically from $\bar{\mu}_0 = B_S(y_0)$.\cite{Johnston2011}  From Eqs.~(\ref{Eq:ChiPerp}) and~(\ref{Eq:ChiParaColl}) one obtains
\be
\frac{\chi_\parallel(T)}{\chi(T_{\rm N})} = \frac{1-f}{\tau^\ast - f}.
\label{Eq:chiTTNColl}
\ee
By Taylor expanding $B_S^\prime(y_0) = (S+1)/3$ for $y_0\to0$, one obtains $\tau^\ast(t\to 1) = 1$ and $\frac{\chi_\parallel(T\to T_{\rm N})}{\chi(T_{\rm N})}=1$, as required.  For $T\to0$, $B_S^\prime(y_0)\to0$, $\tau^\ast\to\infty$ and $\chi_\parallel\to0$.  The  parameters in Eq.~(\ref{Eq:chiTTNColl}) required to fit experimental \mbox{$\chi_\parallel(T\leq T_{\rm N})$} data are just $f,\ S$ and $\chi(T_{\rm N})$, which can usually be easily independently determined from experiment or estimated.  Setting $f=-1$ in Eq.~(\ref{Eq:ChiParaColl}) reproduces Van~Vleck's 1941 prediction for the special case of bipartite collinear AFMs with only nearest-neighbor interactions.\cite{VanVleck1941}

For planar noncollinear AFMs, one must take into account via MFT the field-induced changes in both the magnitudes and directions of the ordered moments to first order in $H$, and we then obtain the in-plane ($xy$) susceptibility
\be
\frac{\chi_{xy}(T\leq T_{\rm N})}{\chi(T_{\rm N})} = \frac{(1+\tau^\ast + 2f + 4B^\ast)(1-f)}{2\left[(\tau^\ast + B^\ast)(1+B^\ast) - (f+B^\ast)^2\right]},
\label{Eq:ChiFinalRatio}
\ee
where
\be
B^\ast = -\frac{\sum_j {J}_{ij}\cos^2\phi_{ji}}{\sum_j {J}_{ij}\cos\phi_{ji}}.
\label{Eq:tauastBast}
\ee
Using $\tau^\ast(t\to 1) = 1$,  Eq.~(\ref{Eq:ChiFinalRatio}) gives $\frac{\chi_{xy}(T\to T_{\rm N})}{\chi(T_{\rm N})} = 1$, irrespective of the value of $B^\ast$, as required, whereas $\lim_{t\to0}B_S^\prime(y_0)\to 0$ and $\tau^\ast\to\infty$ yield from Eq.~(\ref{Eq:ChiFinalRatio})
\be
\frac{\chi_{xy}(T=0)}{\chi(T_{\rm N})} = \frac{1-f}{2(1+B^\ast)}.
\label{Eq:ChiT0Ratio}
\ee
The parameter $B^\ast$ is the only new parameter specifically associated with noncollinear AFMs, is not generally directly  measurable, but can be evaluated if the AFM structure and an exchange interaction model are available.  Alternatively, it can be used as a fitting parameter to provide such information.

On the other hand, the value of $B^\ast$ can be experimentally determined within a minimal generic $J_0$-$J_{z1}$-$J_{z2}$ model\cite{Nagamiya1967} for helical/cycloidal AFM structures as in Fig.~\ref{Fig:J0_Jz1_Jz2_model_helix} on any Bravais spin lattice.  In this model, one sums the exchange interactions of a given magnetic moment with all other moments in the same ferromagnetically-aligned layer perpendicular to the helical/cycloidal wave vector {\bf k}  and calls that sum $J_0$, and similarly for nearest- and next-nearest-layer  interactions $J_{z1}$ and $J_{z2}$, respectively, as indicated in Fig.~\ref{Fig:J0_Jz1_Jz2_model_helix}.  The same theory is applicable to isolated linear chains where $J_0=0$.  Then the {\bf k} of the helix/cycloid  is obtained in terms of the exchange constants by minimizing the exchange energy to be\cite{Yoshimori1959, Nagamiya1967}
\be
\cos(kd) = -\frac{J_{z1}}{4J_{z2}},
\label{Eq:coskzd}
\ee
where $k=|{\bf k}|$ and $d$ is the distance between layers.  $kd$ is the turn angle between adjacent moments along the helix/cycloid axis (Fig.~\ref{Fig:J0_Jz1_Jz2_model_helix}) and is experimentally measurable by magnetic x-ray or neutron diffraction techniques.  Using Eq.~(\ref{Eq:coskzd}) one can express $B^\ast$ in Eq.~(\ref{Eq:tauastBast}) as
\be
B^\ast = 2(1-f)\cos(kd)[1+\cos(kd)] - f.
\label{Eq:BastFromfkz}
\ee
Using Eq.~(\ref{Eq:BastFromfkz}), one can now write $\chi_{xy}(T\leq T_{\rm N})/\chi(T_{\rm N})$ in Eq.~(\ref{Eq:ChiFinalRatio}) completely in terms of independently measurable quantities.  Furthermore, using Eqs.~(\ref{Eq:ChiT0Ratio}) and~(\ref{Eq:BastFromfkz}) one obtains
\be
\frac{\chi_{xy}(T=0)}{\chi(T_{\rm N})} = \frac{1}{2\big[1+2\cos (kd) + 2\cos^2(kd)\big]}.
\label{Eq:ChiT0TNkz}
\ee
The expression for $\chi_{xy}(T=0)/\chi(T_{\rm N})$ obtained in 1959 by Yoshimori\cite{Yoshimori1959} for the special case of the helix in MnO$_2$ is consistent with the general result~(\ref{Eq:ChiT0TNkz}).

\begin{figure}
\includegraphics [width=2.5in]{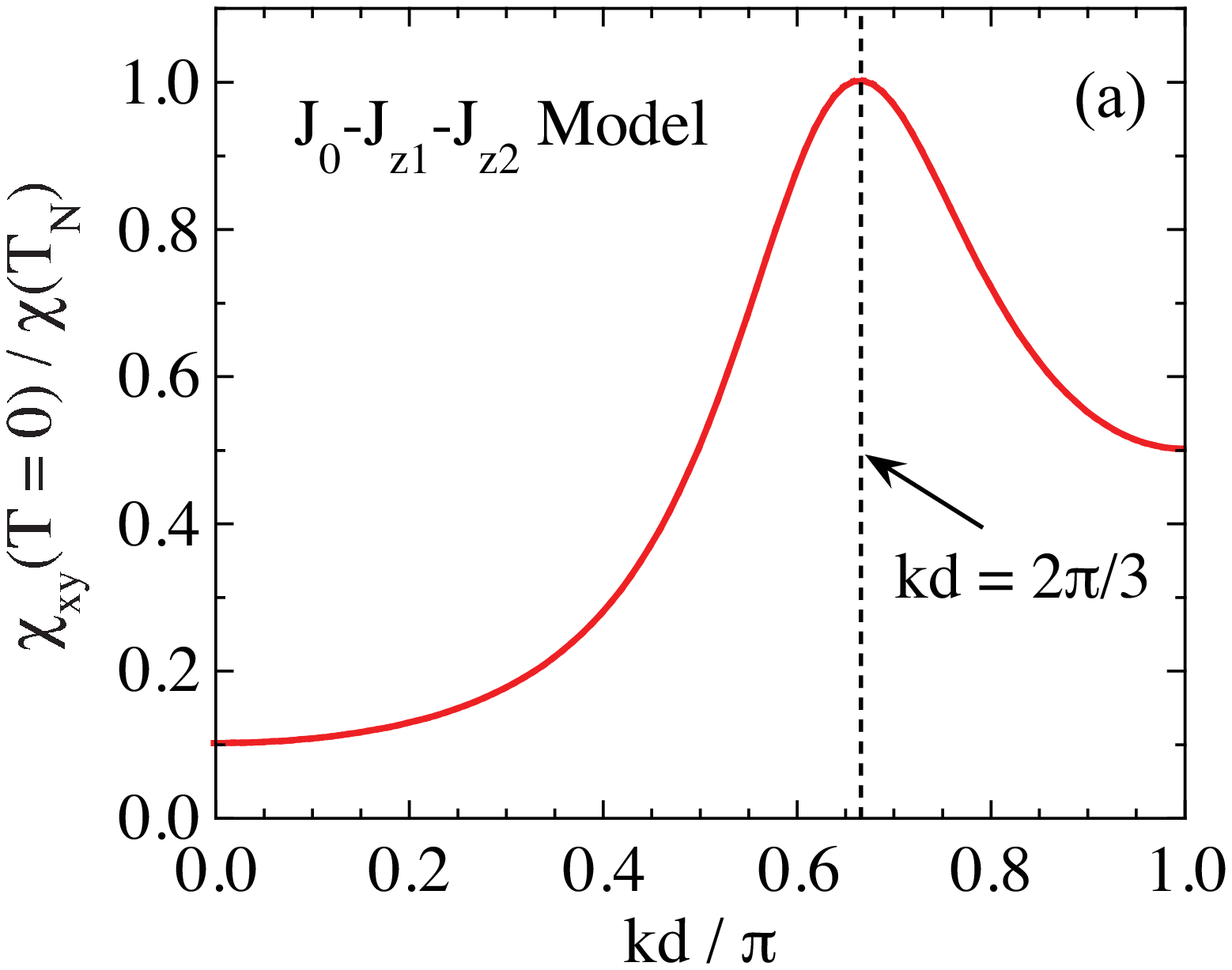}
\includegraphics [width=2.5in]{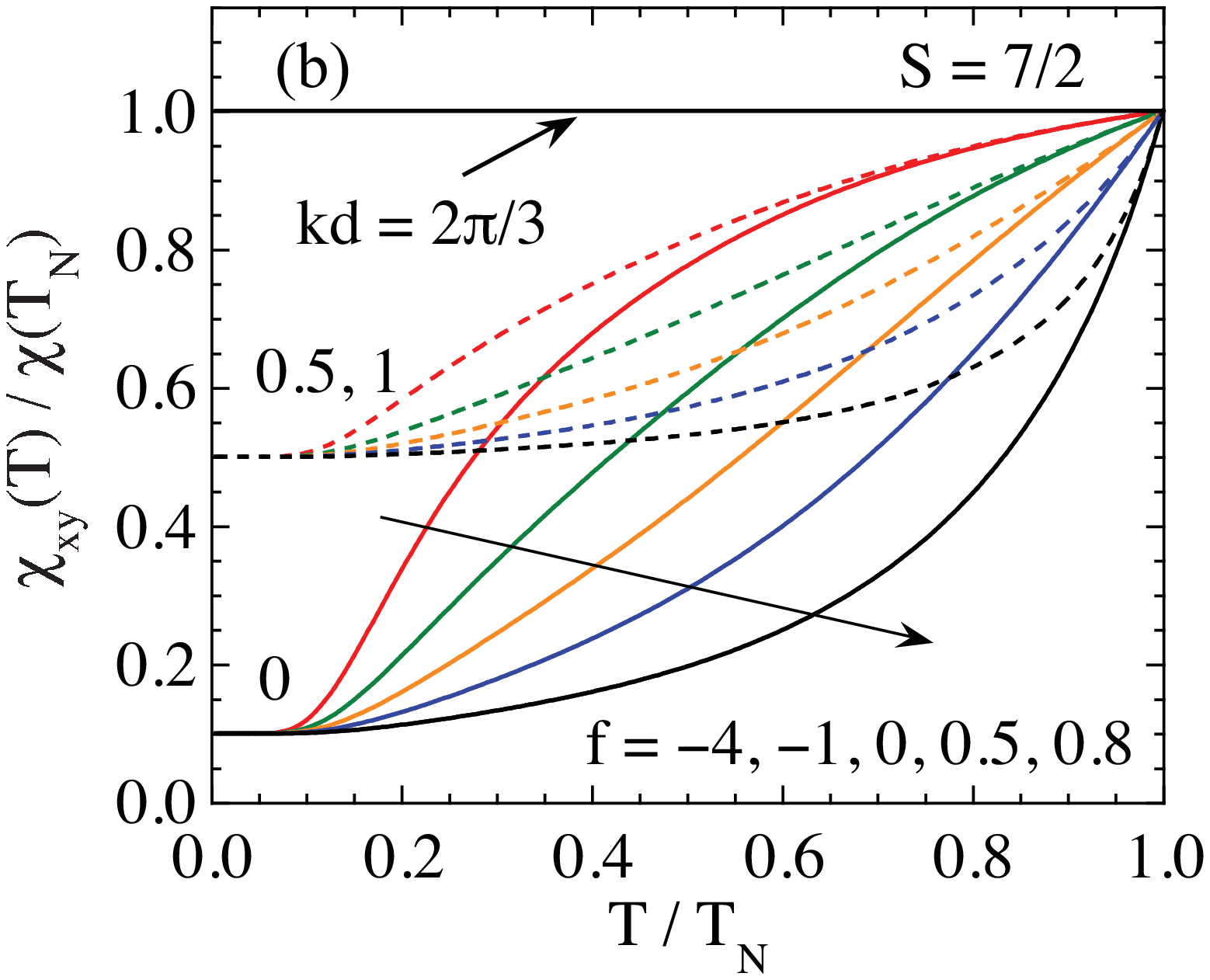}
\caption{(Color online) (a) $\chi_{xy}(T=0)$ versus $kd$ [Eq.~(\ref{Eq:ChiT0TNkz})] and (b) $\chi_{xy}$ versus $T$ and $f$ [Eqs.~(\ref{Eq:ChiFinalRatio}), (\ref{Eq:BastFromfkz})] for helical/cycloidal AFMs with $S = 7/2$ within the generic $J_0$-$J_{z1}$-$J_{z2}$ MFT model. }
\label{Fig:J0Jz1Jz2Chi0onChiTN}
\end{figure}

\begin{figure}
\includegraphics [width=2.5in]{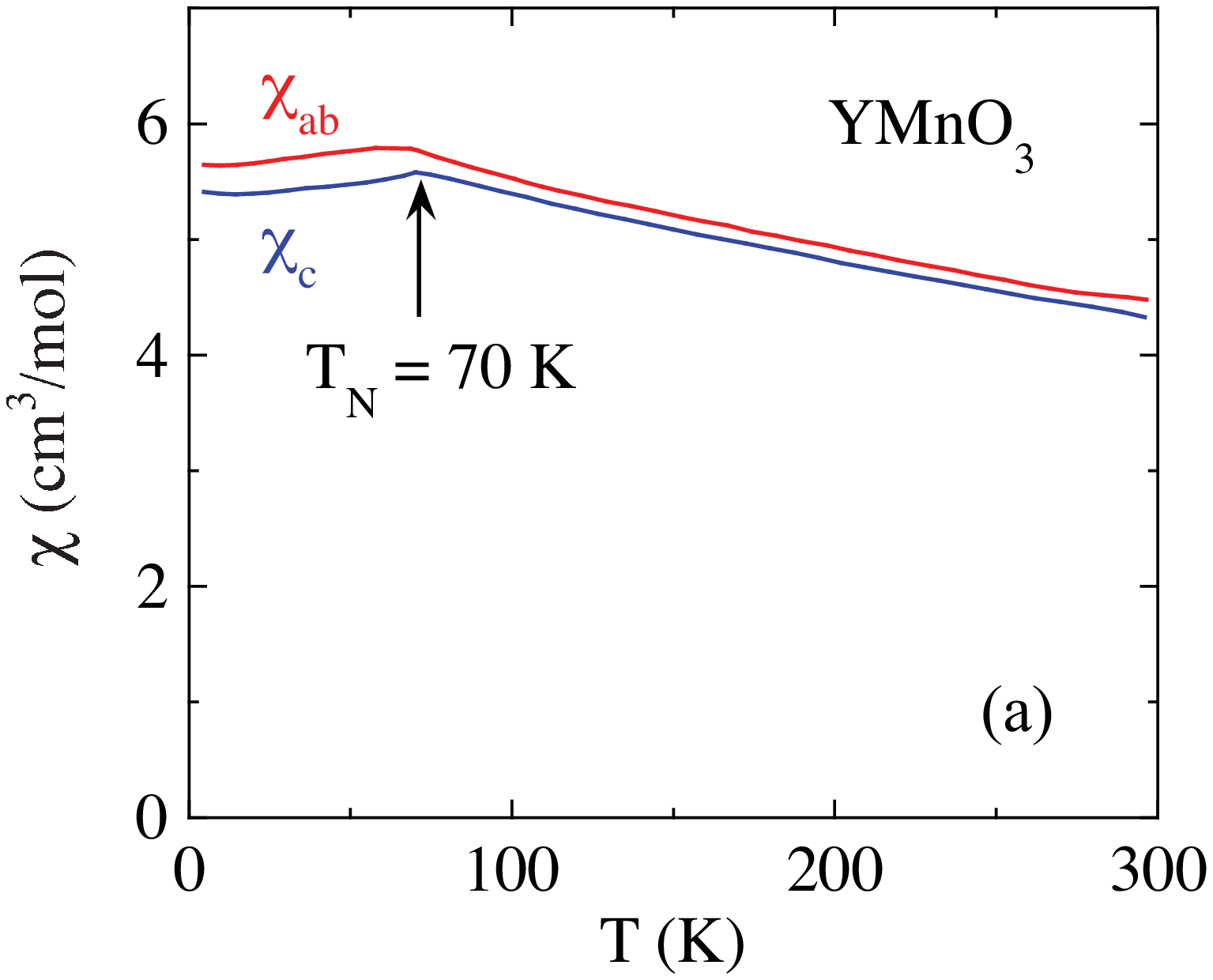}
\includegraphics [width=2.45in]{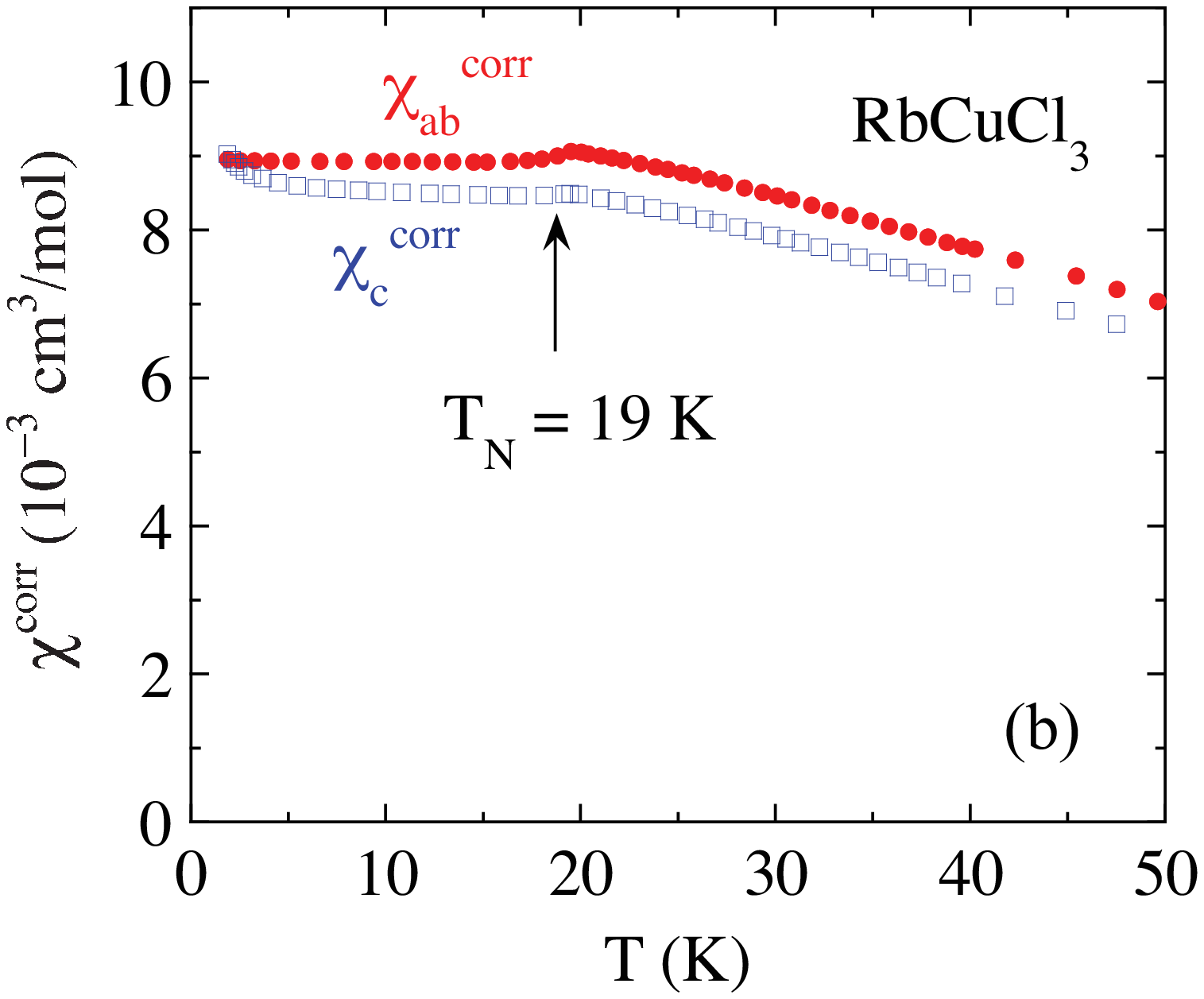}
\caption{(Color online) Anisotropic susceptibilities $\chi_{ab}(T)$ for the triangular lattice AFMs (a) hexagonal ${\rm YMnO_3}$ (Ref.~\onlinecite{Katsufuji2001})  with Mn$^{+3}$ spins $S = 2$ and $kd=120^\circ$ ordering in the $ab$-plane\cite{Brown2006} and (b) RbCuCl$_3$ with Cu$^{+2}$ spins $S = 1/2$ (Ref.~\onlinecite{Maruyama2001}) that has a slightly distorted triangular spin lattice with $kd=108^\circ$ cycloidal ordering in the hexagonal $ab$-plane.\cite{Reehuis2001}  The $\chi(T)$ data in (b) are corrected for the $T$-independent diamagnetic core and paramagnetic Van~Vleck contributions.\cite{Maruyama2001} }
\label{Fig:Triangular_Lattice_AF_k}
\end{figure}

Using Eq.~(\ref{Eq:ChiT0TNkz}), the $\chi_{xy}(T=0)/\chi(T_{\rm N})$ is plotted versus $kd$ in Fig.~\ref{Fig:J0Jz1Jz2Chi0onChiTN}(a).  The predicted behavior has a surprising nonmonotonic dependence on $kd$ with a maximum at $kd= 2\pi/3$ with a value of unity.   Using Eqs.~(\ref{Eq:ChiFinalRatio}) and~(\ref{Eq:BastFromfkz}), $\chi_{xy}(T\leq T_{\rm N})$ and its dependences on $kd$ and $f$ are shown in Fig.~\ref{Fig:J0Jz1Jz2Chi0onChiTN}(b), where $\chi_{xy}$ is seen to be strongly dependent on $T$ and $f$ except for $kd=2\pi/3 = 120^\circ$ for which it is independent of $T$ and $f$.  One can prove that this result for $kd=2\pi/3$ is obtained within the $J_0$-$J_{z1}$-$J_{z2}$ model for any value of $S$.  Then using Eq.~(\ref{Eq:ChiPerp}), our MFT makes the remarkable universal prediction for helical/cycloidal 120$^\circ$ AFM ordering that $\chi(T\leq T_{\rm N})/\chi(T_{\rm N})$ is isotropic and independent of $S$, $f$ and $T$ for $T\leq T_{\rm N}$.  The same result is obtained for other AFMs with 120$^\circ$ ordering and therefore a helical/cycloidal AFM structure is not required [see also Fig.~\ref{Fig:Triangular_Lattice_AF_k}(a) below].

If only the six nearest-neighbor interactions $J$ occur in a single triangular lattice layer exhibiting 120$^\circ$ ordering in MFT, one obtains from Eqs.~(\ref{Eq:CThetap}) and~(\ref{Eq:ChiPerp}) that \mbox{$\chi(T=0)/(Ng^2\mu_{\rm B}^2) = 1/(9J)$}, independent of $S$\@.  For the classical ($S\to\infty$) isolated triangular layer Heisenberg AFM, one obtains the same value.\cite{Kawamura1985, Chubukov1994}  Classical Monte Carlo simulations for a triangular spin lattice layer indicate that $\chi$ is isotropic and also nearly independent of $T$ at low $T$.\cite{Kawamura1984}  Our MFT result for $kd= 2\pi/3$ thus significantly extends the previous calculations for single classical triangular lattice layers to finite quantum spins $S$ and long-range AFM ordering of coupled layers.

\begin{figure}
\includegraphics [width=2.5in]{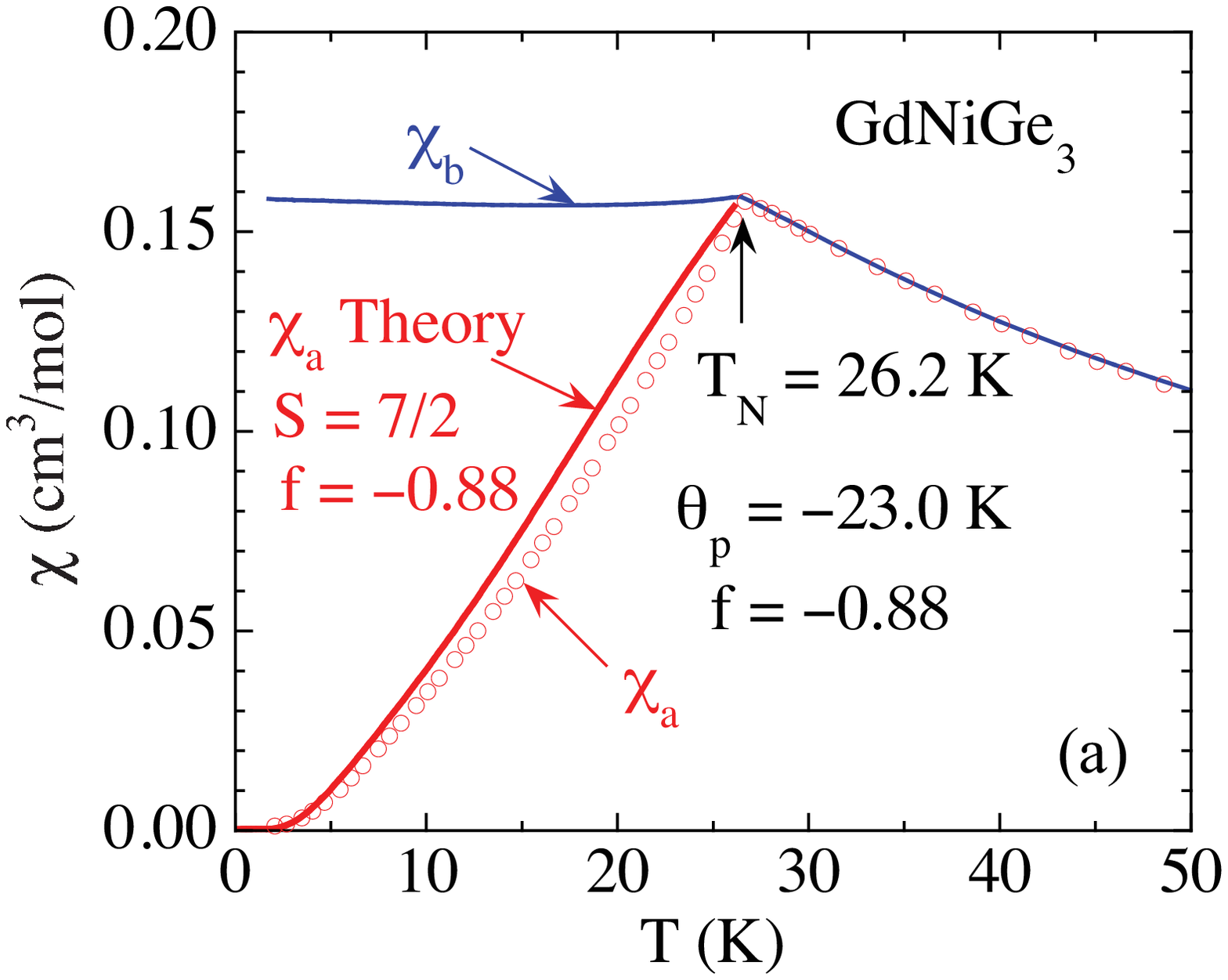}
\includegraphics [width=2.5in]{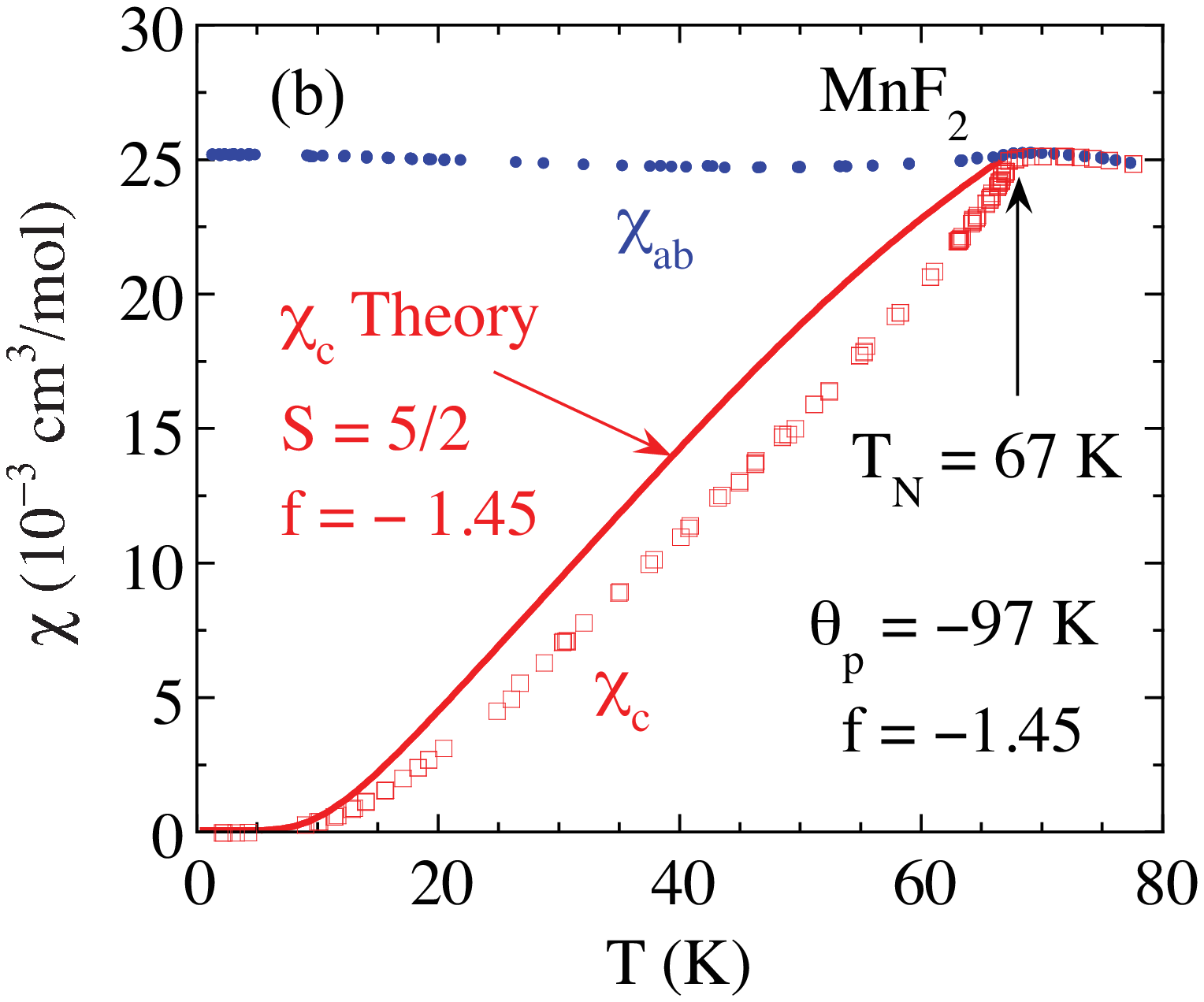}
\caption{(Color online) Anisotropic $\chi(T)$ of single crystals of (a) GdNiGe$_3$ (Ref.~\onlinecite{Mun2010}) and (b) MnF$_2$.\cite{Trapp1963, Trapp1963a} In (a), the easy-axis is the $a$-axis (open circles), whereas in (b) it is the $c$-axis.  In (b), the $\theta_{\rm p}$ value was taken from Ref.~\onlinecite{Corliss1950}.  The corresponding $\chi_\parallel(T)$ data are fitted by the MFT prediction in Eq.~(\ref{Eq:chiTTNColl}) (solid red curves) with no adjustable parameters.}
\label{Fig:GdNiGe3_MnF2_chi}
\end{figure}

\begin{figure}
\includegraphics [width=2.in]{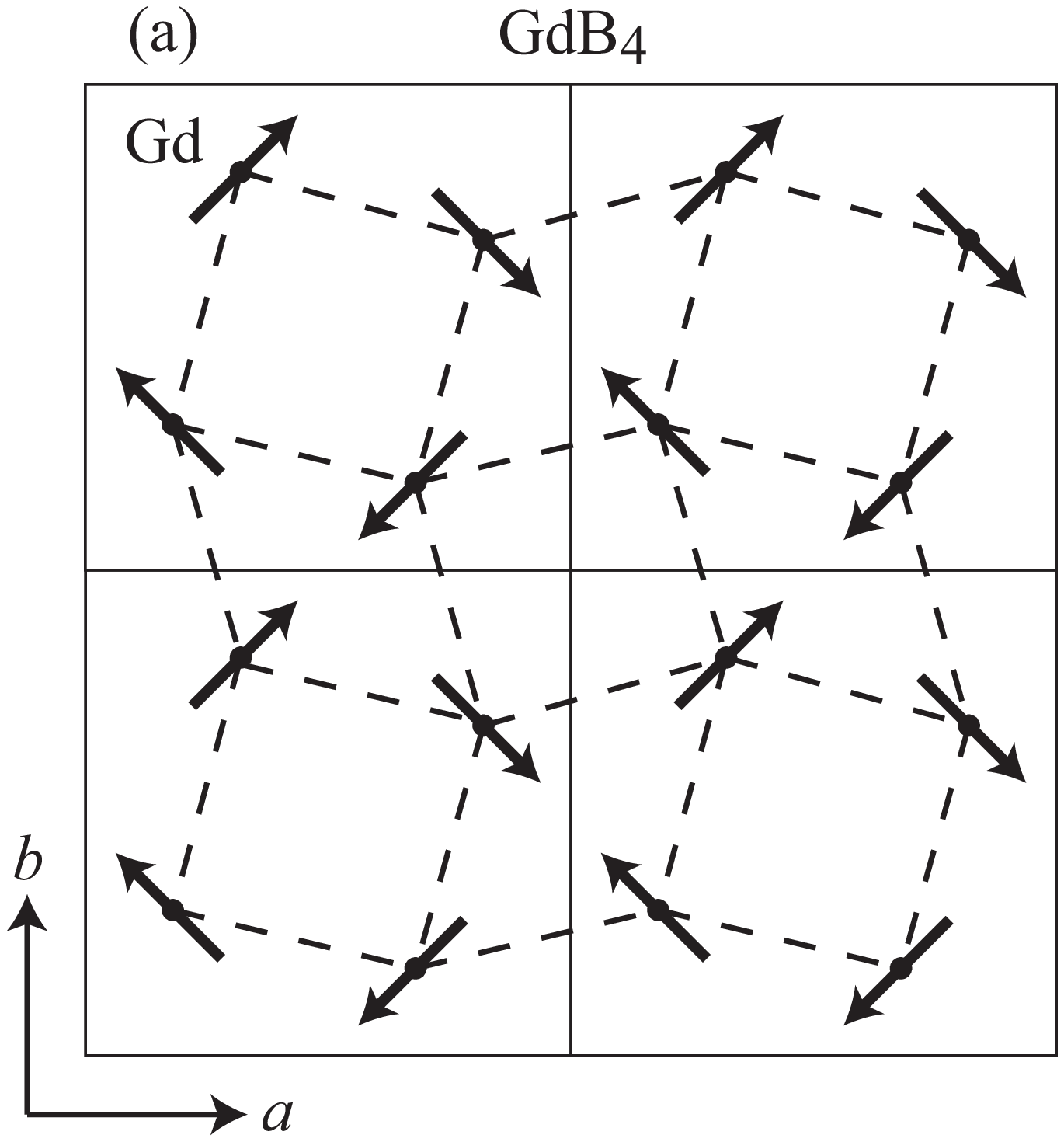}
\includegraphics [width=2.5in]{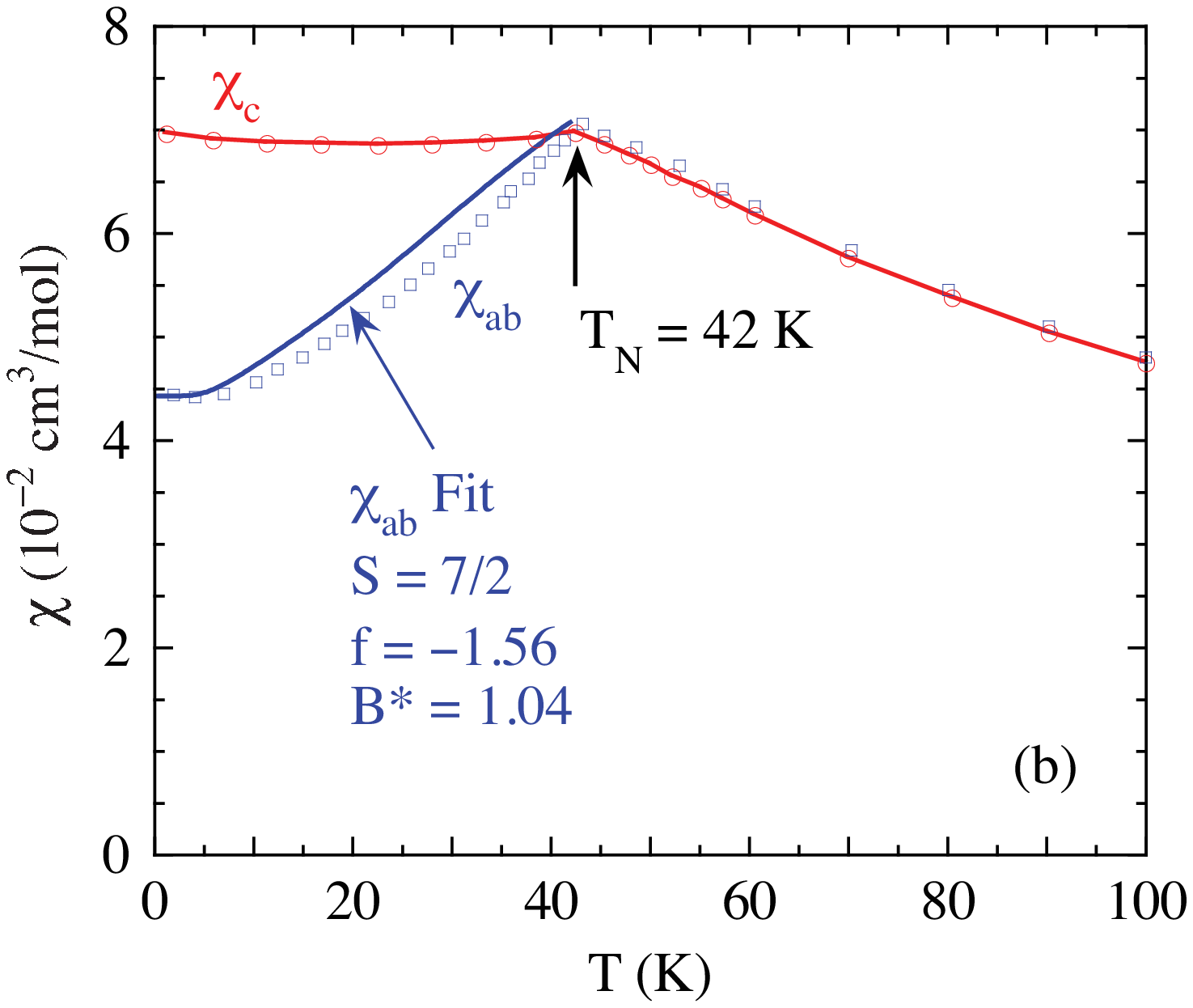}
\caption{(Color online) (a) AFM structure of GdB$_4$.\cite{Brown2006}  (b) Magnetic susceptibility $\chi$ versus temperature $T$ for a single crystal of GdB$_4$,\cite{Cho2005} together with the fit of the $\chi_{ab}(T\leq T_{\rm N})$ data by Eq.~(\ref{Eq:ChiFinalRatio}) using $S=7/2$ and the experimentally determined parameters $f$ and $B^\ast$ in the figure.  Thus there are no adjustable parameters in the fit.}
\label{Fig:GdB4_Cho_chi}
\end{figure}

\paragraph*{Fits of Experimental Data.} As shown in Eq.~(\ref{Eq:ChiPerp}), $\chi_\perp$ is independent of $T$ below $T_{\rm N}$ with the value $\chi(T_{\rm N})$, so no explicit fitting of experimental data is required.

We first present fits by Eq.~(\ref{Eq:chiTTNColl}) of $\chi_\parallel(T)$ data for the collinear AFMs GdNiGe$_3$, an orthorhombic compound containing nonmagnetic Ni atoms and Gd$^{+3}$ spins $S = 7/2$,\cite{Mun2010} and MnF$_2$ with the primitive tetragonal rutile structure containing Mn$^{+2}$ spins $S = 5/2$.\cite{Stout1954}  The anisotropic $\chi(T)$ data at low~$T$ for single crystals of GdNiGe$_3$ (Ref.~\onlinecite{Mun2010}) and MnF$_2$ (Refs.~\onlinecite{Trapp1963, Trapp1963a}) and the corresponding fits of the $\chi_\parallel(T\leq T_{\rm N})$ data by Eq.~(\ref{Eq:chiTTNColl}) with no adjustable parameters are shown in Fig.~\ref{Fig:GdNiGe3_MnF2_chi}.  The fit to the $\chi_\parallel(T\leq T_{\rm N})$ $a$-axis data of GdNiGe$_3$ with $S = 7/2$  is better than the fit to the corresponding $c$-axis data of MnF$_2$ with $S = 5/2$.  This comparison agrees with expectation, because MFT does not include the influence of quantum spin fluctuations which increase as $S$ decreases.  This suggests that a comparison of such MFT fits with experimental data is a quantitative diagnostic for the occurrence  at $T\leq T_{\rm N}$ of spin fluctuations and correlations beyond MFT\@.

As an example of a noncollinear planar AFM, primitive tetragonal GdB$_4$ consists of crystallographically equivalent Gd spins~$7/2$ with the AFM structure shown in Fig.~\ref{Fig:GdB4_Cho_chi}(a) and with the ordered moments oriented in the [110] and equivalent directions.\cite{Blanco2006}  The magnetic and chemical unit cells are the same.  Anisotropic $\chi(T)$ data at low $T$ are shown in Fig.~\ref{Fig:GdB4_Cho_chi}(b).\cite{Cho2005}  The fit of the \mbox{$\chi_{ab}(T\leq T_{\rm N})$} data by Eq.~(\ref{Eq:ChiFinalRatio}) with no adjustable parameters is shown by the solid blue curve using parameters in the figure.  The value of $B^\ast$ was estimated from Eq.~(\ref{Eq:ChiT0Ratio}) and the experimental values\cite{Cho2005} of $f$ and $\chi_{ab}(T\to0)/\chi(T_{\rm N})$.  The relationship between the fit and data is similar to that for GdNiGe$_3$ in Fig.~\ref{Fig:GdNiGe3_MnF2_chi}(a).

We now test our universal prediction for noncollinear 120$^\circ$ AFM structures that $\chi(T\leq T_{\rm N})$ is isotropic and independent of $f,\ S$ and $T$ for $0\leq T\leq T_{\rm N}$ with the value $\chi(T_{\rm N})$, which does not require explicit fits.  The hexagonal compound $\alpha$-${\rm YMnO_3}$ contains a triangular lattice of crystallographically equivalent Mn$^{+3}$ spins $S=2$ and exhibits $120^\circ$ coplanar ordering in the $ab$-plane.\cite{Brown2006}  As in GdB$_4$, the magnetic and chemical unit cells are the same.  Anisotropic $\chi(T)$ data for this compound are shown in Fig.~\ref{Fig:Triangular_Lattice_AF_k}(a).\cite{Katsufuji2001}   The $\chi(T\leq T_{\rm N})$ data parallel and perpendicular to the $ab$-plane are nearly isotropic and independent of $T$\@.  Similar $\chi(T\leq T_{\rm N})$ results have been obtained for many triangular lattice AFMs with $120^\circ$ helical or cycloidal ordering, such as the $S = 3/2$ compounds LiCrO$_2$,\cite{Kadowaki1995} VF$_2$ and VBr$_2$.\cite{Hirakawa1983, Kadowaki1985, Kadowaki1987}  Our MFT prediction is even strongly confirmed by the $\chi(T\leq T_{\rm N})$ data\cite{Maruyama2001} in Fig.~\ref{Fig:Triangular_Lattice_AF_k}(b) for the slightly monoclinically distorted triangular spin lattice in RbCuCl$_3$ containing highly quantum Cu$^{+2}$ spins-1/2 exhibiting cycloidal AFM ordering within the hexagonal $ab$-plane.\cite{Reehuis2001}  The cycloid axis is in the hexagonal [110] direction with a turn angle $kd = 108^\circ$,\cite{Reehuis2001} close to the undistorted triangular lattice value of 120$^\circ$.  The reason that the MFT prediction is accurate even for $S = 1/2$ deserves further investigation.

In summary, a generic molecular field theory of the anisotropic \mbox{$\chi(T\leq T_{\rm N})$} was formulated for local moment Heisenberg AFMs that is widely applicable to collinear and planar noncollinear AFM structures.  The comparisons of our results with experimental anisotropic $\chi(T\leq T_{\rm N})$ data for single crystals in Figs.~\ref{Fig:GdNiGe3_MnF2_chi}--\ref{Fig:Triangular_Lattice_AF_k} with no adjustable parameters demonstrate that such analyses constitute a powerful probe of the AFM structure and spin interactions.  Our results will also be useful for analyzing $\chi(T \leq T_{\rm N})$ data for polycrystalline samples.  An important avenue for future research is to further study the applicability, accuracy and limitations of our MFT predictions.  The present work is a stepping stone for additional MFT calculations of $\chi(T \leq T_{\rm N})$ that could include various types of anisotropies.

\acknowledgments

The author is grateful to \mbox{A.~Honecker} and M.~E.~Zhitomirsky for insights about the $\chi$ of triangular lattice AFMs, and to \mbox{S. L. Bud'ko} and \mbox{H. Tanaka} for communicating $\chi(T)$ data.  This research at Ames Laboratory was supported by the U.S. Department of Energy, Office of Basic Energy Sciences under Contract No.~DE-AC02-07CH11358.

\end{document}